\documentclass[11pt,a4paper]{article}
\usepackage{jcappub}

\usepackage{amsmath}
\usepackage{amsfonts}
\usepackage{amsthm}
\def\be{\begin{equation}}
\def\ee{\end{equation}}
\def\bea{\begin{eqnarray}}
\def\eea{\end{eqnarray}}

\newcommand{\f}{\Sigma}
\newcommand{\V}{V}
\newcommand{\U}{U}

\newcommand{\trho}{\varphi}
\newcommand{\brho}{\rho}
\newcommand{\prho}{\rho^\prime}
\begin{document}

\title{String Fluid in Local Equilibrium}
\author{Daniel Schubring and}

\emailAdd{schub071@d.umn.edu}

\author{Vitaly Vanchurin}

\emailAdd{vvanchur@d.umn.edu}

\date{\today}

\affiliation{Department of Physics, University of Minnesota, Duluth, Minnesota, 55812}

\abstract{

We study the solutions of string fluid equations under assumption of a local equilibrium which was previously obtained in the context of the kinetic theory. We show that the fluid can be foliated into non-interacting submanifolds whose equations of motion are exactly that of the wiggly strings considered previously by Vilenkin and Carter. In a special case of negligible statistical variance in either the left or the right-moving directions of microscopic strings, the submanifolds are described by the action of chiral strings proposed by Witten. When both variances vanish the submanifolds are described by the Nambu-Goto action and the string fluid reduces to the string dust introduced by Stachel. }
\maketitle

\tableofcontents
\section{Introduction}

A fluid mechanical description of physical systems with many degrees of freedom is often employed whenever local interactions tend to quickly drive the local sub-systems towards equilibrium. Although the true local equilibrium is never established, the perturbative expansion around equilibrium states provides a useful insight into the behavior of the systems as a whole. The key idea of the fluid description is to impose the microscopic conservation laws, such as conservation of energy and momentum, to derive macroscopic equations of motion, such as the continuity and Navier-Stokes equations. 

The fluid approach has proven to be useful for describing many different systems on a wide range of scales in which the ``microscopic'' degrees of freedom are ``zero-dimensional'' particles (e.g. molecules, stars, galaxies), but it remains unclear whether similar ideas can be applied to study ``one-dimensional'' strings (e.g. cosmic strings \cite{Cosmic}, fundamental strings \cite{Fundamental}, topological strings \cite{Crystal}, polymer molecules \cite{Polymer}, etc.). This question goes back to the earlier attempts to develop a kinetic theory  \cite{LoweThorlacius, BarrabesIsrael} and fluid mechanics \cite{StachelDust, CarterBrane, Kopczynski} of strings. In what follows we briefly review our contribution to the field. For more details the reader is referred to the original publications \cite{OldKinetic, Kinetic, Fluid, Transport}.

To formulate a kinetic theory of strings we considered the dynamics of a distribution function of the energy density defined on a space of right- and left-moving null directions $A^\mu$ and $B^\mu$ of the microscopic strings. Under the so-called string chaos assumption it was possible to derive a transport equation for strings similarly to how the molecular chaos assumption is used to derive the Boltzman transport equation for particles \cite{OldKinetic, Kinetic}. The homogeneous transport equation enabled us to prove the $H$-theorem for strings and to solve for the equilibrium distribution \cite{Kinetic}. It was shown that in the equilibrium the right- and left-moving null directions $A^\mu$ and $B^\mu$ are statistically independent. Although it was not immediately clear how to consistently include spatial variations, the correct version of the inhomogeneous transport equation for the Nambu-Goto strings was eventually obtained in \cite{Transport}.

To develop a fluid description of strings we derived conservation equations for a coarse-grained tensor current $ \langle A\otimes B \rangle^{\mu\nu}$. The symmetric part of the equation represents the microscopic conservation of energy and momentum and the antisymmetric part of the equation represents the continuity of individual strings \cite{Fluid}. Although the conservation equations are exact as no assumption were made to derive them, their solutions are not uniquely determined unless additional constraint are imposed. In contrast, the transport equation of the kinetic theory is only approximate, as it relies on the strings chaos assumption, but one can solve it starting from an arbitrary initial condition. This shows that the two approaches are only useful in the ranges of their respective validities, but under assumption of local equilibrium both approaches indeed lead to the same set of fluid equations \cite{Transport}.  

The paper is organized as follows. In Sec. \ref{Sec:Review} we review some basic results for the individual Nambu-Goto strings. In Sec. \ref{Sec:Fluid} we develop a fluid description of Nambu-Goto strings and in Sec. \ref{Sec:Solutions} we analyze different classes of solutions of the fluid equations in the limit of local equilibrium. The main results of the paper are summarized and discussed in Sec. \ref{Sec:Discussion}.

\section{Nambu-Goto Strings} \label{Sec:Review}

We start by reviewing the basic properties of the individual Nambu-Goto strings.  Consider a world-sheet of a single string described by coordinates $ \eta^a$, where $a=0,1$, embedded into the four-dimensional target space $ X^\mu(\eta^a)$, where $\mu=0,1,2,3$. Then we can define a pullback of the target space metric (or the induced metric)
\be
h_{ab} \equiv g_{\mu\nu}X^\mu_{,a}X^\nu_{,b}\label{hmetric}.
\ee
For the Nambu-Goto strings the equations of motions are obtained from the action,
\be
S = -\int d\eta^0\wedge d\eta^1\, \sqrt{-h},\label{action}
\ee
where the units are chosen to set the string tension coefficient to one, and the corresponding (singular) energy-momentum tensor as a function of the target space coordinates $ x^\lambda $ is given by,
\begin{align}
T^{\mu\nu}(x^\lambda)\sqrt{-g(x^\lambda)}=  \int d\eta^0\wedge d\eta^1  \sqrt{-h}\,h^{ab}X^\mu_{,a}X^\nu_{,b}\,  \delta^{(4)}(x^\lambda-X^\lambda).\label{T}
\end{align}
Due to conservations of energy and momenta, the energy-momentum tensor should also obey the conservation equation,
\begin{align}
\nabla_\mu T^{\mu\nu} = 0,\label{cons_T}
\end{align}
but because of the presence of the delta function in \eqref{T}, the interpretation of the expression \eqref{cons_T} is somewhat obscure.

\subsection{Conservation Equations}

To clarify the conservation law \eqref{cons_T} for a singular energy momentum tensor \eqref{T}, consider first a general singular current of the form
\begin{align}
J^\mu(x^\lambda)\sqrt{-g(x^\lambda)} = \int d\eta^0\wedge d\eta^1  \tilde{J}^\mu(\eta)  \,\delta^{(4)}(x^\lambda-X^\lambda)\label{J}.
\end{align}
Then the conserved current $J^\mu$ formally obeys the conservation condition in the target space
\be
\nabla_\mu J^\mu = \frac{1}{\sqrt{-g}} \partial_\mu(J^\mu\sqrt{-g}) = 0
\ee
or
\be
\partial_\mu(J^\mu\sqrt{-g}) = 0.\label{J_conservation}
\ee
By integrating over a four-dimensional volume, this was shown \cite{Fluid} to imply a conservation condition on the worldsheet
\be
\partial_a \tilde{J}^a = 0 \label{singular_conservation}
\ee 
for a  vector $\tilde{J}^a$ which can be pushforward to the conserved current $\tilde{J}^\mu$ in the target space, 
\be
\tilde{J}^\mu = \tilde{J}^a X^\mu_{,a}\label{current_pushforward}.\\
\ee
(See Ref. \cite{Fluid} for details).

The same procedure can be applied directly to the energy-momentum tensor \eqref{T} of a Nambu-Goto string in flat space-time. Then the four  conservation equations \eqref{cons_T} in the target space can be put to the same form as \eqref{J_conservation},
\be
\partial_\mu(T^{\mu\nu} \sqrt{-g}) = 0 \label{current_conservation}
\ee
and by inspecting  \eqref{T} we can we can identify the four conserved currents on the world-sheet as the four coefficients of $ X^\mu_{,a} $ leading to the four familiar equations of motion for Nambu-Goto strings in flat space-time,
\begin{align}
\partial_a (\sqrt{-h}\,h^{ab}X^\nu_{,b}) = 0.\label{NGflat}
\end{align}(For example, the target space current corresponding to $\nu=0$,
\be
\tilde{J}^{\mu}=  \sqrt{-h}\,h^{ab}X^\mu_{,a}X^0_{,b}
\ee
is a push-forward \eqref{current_pushforward} of a world-sheet current $\tilde{J}^a =\sqrt{-h}\,h^{ab}X^0_{,b}$ and thus, according to \eqref{singular_conservation}, the following conservation equation must be satisfied,
\be
\partial_a \tilde{J}^a =\partial_a (\sqrt{-h}\,h^{ab}X^0_{,b})=0,
\ee
which is nothing but equation \eqref{NGflat} for $\nu=0$.)

For a general space-time metric the equivalent of \eqref{current_conservation} is not true for the second rank tensor $ T^{\mu\nu} $ since there is an additional term involving a connection coefficient in the target space conservation equation \eqref{cons_T},
\be
\nabla_\mu  T^{\mu\nu} = \frac{1}{\sqrt{-g}} \partial_\mu (\sqrt{-g} T^{\mu\nu}) + \Gamma^\nu_{\lambda \mu} T^{\lambda \mu} = 0. 
\ee 
But this simply leads to an additional term $ \Gamma^\nu_{\lambda\mu}\tilde{T}^{\lambda\mu} $ in the singular current conservation equation, which can also be pushed-forward to become the Nambu-Goto equation of motion in a general space-time,
\be
X^\mu_{,a}\nabla_\mu (\sqrt{-h}\,h^{ab}X^\nu_{,b}) = 0.\label{T_current}
\ee

Besides the worldsheet currents associated to the energy-momentum tensor, we can consider a trivial current conservation due to commutation of partial derivatives:
\begin{align}
\partial_a \epsilon^{ab}X^\nu_{,b} = 0.\label{F_current}
\end{align}
where $\epsilon^{ab}$ is the Levi-Civita tensor. Following the above discussion, this leads to four more conserved currents 
\begin{align}
\nabla_\mu F^{\mu\nu} = 0 \label{cons_F}
\end{align}
described by a spacetime tensor,
\begin{align}
F^{\mu\nu}(x^\lambda) \sqrt{-g(x^\lambda)}&\equiv \int d\eta^0\wedge d\eta^1 \, \epsilon^{ab}X^\nu_{,b}X^\mu_{,a} \, \delta^{(4)}(x^\lambda-X^\lambda)\label{F}.
\end{align}
The conservation of $F^{\mu\nu}$ is related to the continuity of closed or infinite strings at each point and does not depend on a particular choice of the string action such as the Nambu-Goto action \cite{Fluid}. More generally, in models with open strings (which can have endpoints on monopoles or higher dimensional branes) the conservation equations \eqref{cons_F} may include a source term, but the basic form of the equations would not be expected to change.

\subsection{Right and Left Movers}

In a particular choice of gauge, similarities between $ T^{\mu\nu}$ and $ F^{\mu\nu} $ become apparent. We will denote the two-forms in the integrands of the expressions \eqref{T} and \eqref{F} with a hat,
\begin{align}
\hat{T}^{\mu\nu} &\equiv h^{ab}\,\sqrt{-h}\,X^\mu_{,a}X^\nu_{,b}\,d\eta^0\wedge d\eta^1\label{hatT}\\
\hat{F}^{\mu\nu} &\equiv \epsilon^{ab}X^\mu_{,a}X^\nu_{,b} \,d\eta^0\wedge d\eta^1 = dX^\mu\wedge dX^\nu\label{hatF}
\end{align}
To simplify the factor $ \sqrt{-h}\,h^{ab} $ in \eqref{hatT} we choose $ \eta^{0}$ and $\eta^1$ to be (left-pointing and right-pointing) conformal lightcone coordinates. In this gauge, the equations \eqref{hatT} and \eqref{hatF} become,
\begin{align}
\hat{T}^{\mu\nu}&= 2\mathcal{A}^{(\mu}\mathcal{B}^{\nu)}d\eta^0\wedge d\eta^1\label{That}\\
\hat{F}^{\mu\nu}&= 2\mathcal{A}^{[\mu}\mathcal{B}^{\nu]}d\eta^0\wedge d\eta^1.\label{Fhat}
\end{align}
where the two coordinate basis vectors are denoted as
\bea
\mathcal{A}^\mu\equiv  \frac{\partial X^\mu}{\partial \eta^0}\\
\mathcal{B}^\mu\equiv \frac{\partial X^\mu}{\partial \eta^1} .
\eea

Besides pointing in the two null directions on the worldsheet, $ \mathcal{A}^\mu $ and $ \mathcal{B}^\mu $ are relevant as the two propagation directions of extrinsic perturbations. But it is only the direction which is physically relevant ---there is still some gauge freedom in the normalization. We will define the new vectors $ A^\mu $ and $ B^\mu $ normalized to have a unit time component, i.e.
\begin{align}
A^\mu = \frac{\mathcal{A}^{\mu}}{\mathcal{A}^{0}}\nonumber\\
B^\mu = \frac{\mathcal{B}^{\mu}}{\mathcal{B}^{0}}\label{AB_def}
\end{align}
and expressions \eqref{That} and \eqref{Fhat} can be re-written as,
\begin{align}
\hat{T}^{\mu\nu} = A^{(\mu} B^{\nu)}\, \hat{T}^{00}\nonumber\\
\hat{F}^{\mu\nu} = A^{[\mu} B^{\nu]}\, \hat{T}^{00}\label{TF}.
\end{align}
We can also define the full spacetime tensor,
\begin{align}
(A\otimes B)^{\mu\nu} (x^\lambda) &\equiv T^{\mu\nu} (x^\lambda)+F^{\mu\nu}(x^\lambda)\nonumber\\
&= \frac{1}{\sqrt{-g(x^\lambda)}} \int \hat{T}^{00}\, A^{\mu} B^{\nu}\,  \delta(x^\lambda-X^\lambda(\eta)),\label{AB}
\end{align}
which must satisfy,
\begin{align}
\nabla_\mu(A\otimes B)^{\mu\nu} = \nabla_\nu(A\otimes B)^{\mu\nu} =0 \label{cons_AB},
\end{align}
due to the conservation equations \eqref{cons_T} and \eqref{cons_F}. The string network can also be generalized to contain non-Nambu-Goto strings, and in these cases $ A^\mu$ and $ B^\mu$ will be defined as the physical propagation directions rather than the null directions. In particular, the form of $ \hat{T}^{\mu\nu} $ and $ \hat{F}^{\mu\nu} $ for chiral strings and wiggly strings is identical to the Nambu-Goto case. The only distinction is that one or both of $ A^\mu $ and $ B^\mu $ are timelike vectors rather than null vectors \cite{CarterWiggly,VilenkinWiggly}.

Although the quantities $A^\mu$, $ B^\mu$, and $ T^{00} $ in \eqref{AB} can be defined independently of the choice of gauge, the price we pay is the loss of manifest spacetime covariance. But since $T^{\mu\nu}$, $F^{\mu\nu}$ and, thus, $(A\otimes B)^{\mu\nu}$ all transform as second rank tensors, the null vectors $A^\mu$ and $B^\mu$ are uniquely determined in each frame even if their transformation laws are not those of four-vectors. Instead of $ T^{00} $ it may seem more natural to consider a fully covariant measure such as $T^\mu_{\phantom{\mu}\mu}$. According to \eqref{That}, this is also proportional to the worldsheet area
\begin{align}
\hat{T}^{\mu}_{\,\,\,\mu}=2 \sqrt{-h}\,d\eta_0\wedge\eta_1\label{area}.
\end{align}
But this measure can be recovered from the quantities $A^\mu$, $ B^\mu$, and $ T^{00} $ through \eqref{TF}, and will not be as useful in considering the coarse-grained dynamics.

\section{Fluid Equations}\label{Sec:Fluid}

To develop a fluid description of strings we consider the singular tensor currents $ (A\otimes B)^{\mu\nu} $ of all strings in a local neighborhood around each space-time point, $x^\lambda$. The coarse-grained currents are then determined by integrating the singular currents over a spacetime volume $ \Delta V $about $ x^\lambda$,\footnote{As usual, the fluid approximation relies on the assumption that the coarse-grained fields do not depend significantly on the choice of $ \Delta V $ as long as it is from an appropriate range of scales.} 
\begin{align}
\langle A\otimes B\rangle^{\mu\nu}(x^\lambda) \equiv \frac{1}{\Delta V}\int_{\Delta V} d^4 x \,(A\otimes B)^{\mu\nu}.\label{average}
\end{align}
Using \eqref{AB} the integral in \eqref{average} can be calculated by integrating over different pieces of world-sheets enclosed in the volume $ \Delta V $ with the energy density $ \hat{T}^{00} $ as a measure of integration. Then expectation values of the $A^\mu$ and $B^\mu$ vectors (denoted with a bar) are given by
\begin{align}
\bar{A}^\mu= \frac{1}{\brho}\langle A\otimes B\rangle^{\mu 0}\\
\bar{B}^\nu= \frac{1}{\brho}\langle A\otimes B\rangle^{0 \nu}
\end{align}
where 
 \be
 \brho \equiv \langle A\otimes B\rangle^{00}
 \ee 
is the coarse-grained energy density. 

Since the spatial components of the string network quantities $A^i$ and $B^i$ lie on a unit two-sphere (known as the Kibble-Turok sphere), the variances of the averaged fields $ \bar{A}^\mu$ and $\bar{B}^\nu$ satisfy simple expressions:
\begin{align}
\text{Var}(\bar{A})&=\overline{(A^i A_i)}-\bar{A}^i\bar{A}_i = \bar{A}^\mu\bar{A}_\mu \label{VarA}\\
\text{Var}(\bar{B})&=\bar{B}^\mu\bar{B}_\mu. \label{VarB}
\end{align}
Because of this we will refer to the squares of the
 four-vector magnitudes of $\bar{A}^\mu$ and $\bar{B}^\mu$ as the variances of $ A^\mu$ and $ B^\mu$.

We can now impose the microscopic conservation equations \eqref{cons_AB} to derive macroscopic equations for the coarse-grained field
\be
\nabla_\mu \langle A\otimes B \rangle^{(\mu\nu)} =0 
\label{eq:symmetric}
\ee
and
\be
\nabla_\mu \langle A\otimes B \rangle^{[\mu\nu]}=0.
\label{eq:antisymmetric}
\ee
These equations are generically underdetermined which can be seen by counting the degrees of freedom. A general second rank tensor $ \langle A\otimes B \rangle^{\mu\nu}$ has $16$ independent components, but there are only $4$ dynamical equations in \eqref{eq:symmetric} and $3$ dynamical (corresponding to $\nu =1,2,3$) and $1$ constraint (corresponding to $\nu=0$) equation in \eqref{eq:antisymmetric}. This means that the set of equations can only be solved if we reduce the total number of independent components in  $ \langle A\otimes B \rangle^{\mu\nu}$  to $6$. 

To constrain the underdetermined conservation equations \eqref{eq:symmetric} and \eqref{eq:antisymmetric} we will use the further assumption that $A^\mu$ and $B^\mu$ are statistically independent under the energy-density measure of integration as in equation \eqref{average}. Earlier work on a kinetic theory for string networks indicates that under certain conditions the measure will indeed converge to an equilibrium distribution in which $ A^\mu$ and $ B^\mu$ are independent random variables \cite{Kinetic}. Throughout paper we will adopt this \emph{local equilibrium} assumption under which
\be
\langle A\otimes B\rangle^{\mu\nu} = \brho \bar{A}^\mu \bar{B}^\nu \label{equilibrium}
\ee
and in the last section we will comment on a possible generalization of the string fluid to include the effects of pressure and viscosity which are expected to be important for the fluids of, for example, cosmic strings. 
In the equilibrium fluid the coarse-grained tensors \eqref{TF} become
\begin{align}
\langle T \rangle^{\mu\nu} = \brho \bar{A}^{(\mu}\bar{B}^{\nu)}\\
\langle F \rangle^{\mu\nu} = \brho \bar{A}^{[\mu}\bar{B}^{\nu]},
\end{align}
and the conservation equations \eqref{eq:symmetric} and \eqref{eq:antisymmetric} are greatly simplified \cite{Fluid},
\begin{align}
\nabla_\mu (\brho \bar{A}^\mu\bar{B}^\nu) = 0\label{AB_mu}\\
\nabla_\nu (\brho \bar{A}^\mu\bar{B}^\nu) = 0\label{AB_nu}.
\end{align}
Then the number of degrees of freedom is exactly $6$ described by the components of the three-vectors $A^i$ and $B^i$. As we shall argue below the corresponding equations for  $A^i$ and $B^i$ are completely decoupled from the equations for the energy density, $\rho$ which is no longer an independent degree of freedom.  Once the space-time solutions for $A^i$ and $B^i$ are obtained, the energy density $\rho$ is uniquely determined from certain boundary conditions.

\subsection{Submanifold Structure}

As was already mentioned in the last section, the full tensor $(A\otimes B)^{\mu\nu}$ is a covariant second rank tensor, but $A^\mu$ and $B^\mu$ do not transform as four-vectors under general coordinate transformations. Similarly, the coarse-grained tensor $ \langle A\otimes B \rangle^{\mu\nu} $ is covariant but the individual quantities $ \brho $, $ \bar{A}^\mu $ and $ \bar{B}^\mu $ appear to depend on the coarse-graining frame. It is valid to simply take these quantities to transform covariantly, but then in a transformed frame they will no longer have a simple interpretation as coarse-grained quantities. For instance, if we take $ \brho $ to transform as a scalar, in a new frame it will no longer equal to the energy density, which transforms as a component of a tensor. For the moment, we will take this approach. Later on we will renormalize these quantities in a more manifestly covariant way.

Given these considerations, it is valid to use the product rule to expand \eqref{AB_mu}:
\begin{align}
\nabla_\mu (\brho \bar{A}^\mu\bar{B}^\nu) = \bar{B}^\nu\nabla_\mu (\brho \bar{A}^\mu) + \brho \bar{A}^\mu\partial_\mu \bar{B}^\nu + \brho \bar{A}^\mu \Gamma^{\nu}_{\mu\lambda}\bar{B}^\lambda = 0\label{fluideq_1}
\end{align}
but since $\bar{A}^0=\bar{B}^0 = 1 $, the $ \nu = 0 $ component of equation \eqref{fluideq_1} leads to,
\begin{align}
\nabla_\mu (\brho \bar{A}^\mu) = - \brho  \Gamma^{0}_{\mu\lambda}\bar{A}^\mu\bar{B}^\lambda.\label{fluideq_2}
\end{align}
and by substituting \eqref{fluideq_2} back into \eqref{fluideq_1},
\begin{align}
\bar{A}^\mu\partial_\mu \bar{B}^\nu &= -\Gamma^\nu_{\mu\lambda} \bar{A}^\mu \bar{B}^\lambda +\Gamma^0_{\mu\lambda} \bar{A}^\mu \bar{B}^\lambda \bar{B}^\nu.\label{fluideq_3}
\end{align}
Similarly beginning from \eqref{AB_nu} we get,
\begin{align}
\nabla_\mu (\brho \bar{B}^\mu) = - \brho  \Gamma^{0}_{\mu\lambda}\bar{B}^\mu\bar{A}^\lambda.\label{fluideq_4}
\end{align}
and
\begin{align}
\bar{B}^\mu\partial_\mu \bar{A}^\nu &= -\Gamma^\nu_{\mu\lambda} \bar{B}^\mu \bar{A}^\lambda +\Gamma^0_{\mu\lambda} \bar{B}^\mu \bar{A}^\lambda \bar{A}^\nu.\label{fluideq_5}
\end{align}
In total we get the four equations \eqref{fluideq_2},\eqref{fluideq_3},\eqref{fluideq_4}, and \eqref{fluideq_5} which can be written as
\begin{align}
\nabla_\mu (\brho \bar{A}^\mu) &= - \brho  \,\Gamma^{0}_{\kappa\lambda}\bar{A}^\kappa\bar{B}^\lambda\label{fluideq_rhoA}\\
\nabla_\mu (\brho \bar{B}^\mu) &= - \brho  \,\Gamma^{0}_{\kappa\lambda}\bar{A}^\kappa\bar{B}^\lambda\label{fluideq_rhoB}\\
\bar{A}^\mu\nabla_\mu \bar{B}^\nu &= \Gamma^0_{\kappa\lambda} \bar{A}^\kappa \bar{B}^\lambda \,\bar{B}^\nu\label{fluideq_ADB}\\
\bar{B}^\mu\nabla_\mu \bar{A}^\nu &= \Gamma^0_{\kappa\lambda} \bar{A}^\kappa \bar{B}^\lambda \,\bar{A}^\nu\label{fluideq_BDA}.
\end{align}
In particular equations \eqref{fluideq_ADB} and  \eqref{fluideq_BDA} imply that the commutator  
\begin{align}
[\bar{A},\bar{B}]^\nu = \bar{A}^\mu\nabla_\mu \bar{B}^\nu -  \bar{B}^\mu\nabla_\mu \bar{A}^\nu = \Gamma^0_{\kappa\lambda} \bar{A}^\kappa \bar{B}^\lambda \,(\bar{B}^\nu - \bar{A}^\nu)
\end{align}
lies everywhere in the space spanned by $  \bar{A}^\mu $ and $ \bar{B}^\mu $. Thus by Frobenius' theorem, space-time can be foliated by a family of two-dimensional submanifolds everywhere tangent to $\bar{A}^\mu $ and $ \bar{B}^\mu$. These submanifolds may be thought of as the worldsheets of the one-dimensional field lines of the spacelike vector field $ \bar{B}^\mu - \bar{A}^\mu $, which is nothing but the vector field describing the average local direction (or tangent vector) of strings. 

These submanifolds clarify the Cauchy problem for the string fluid in local equilibrium. If $ \bar{A}^\mu $ and $ \bar{B}^\mu $ are specified on a field line at an initial time, equations \eqref{fluideq_ADB} and \eqref{fluideq_BDA} can be used to solve for the values of $\bar{A}^\mu $ and $ \bar{B}^\mu$ along the full submanifold. The possibility of the intersection of submanifolds physically indicates shockwaves which are not resolved in the equilibrium fluid \cite{Fluid}. But if $\bar{A}^\mu $ and $ \bar{B}^\mu$ are given as initial conditions then the solution can be propagated forward for at least some finite time. Notice that the solution of equations \eqref{fluideq_ADB} and \eqref{fluideq_BDA} for $\bar{A}^\mu $ and $ \bar{B}^\mu$ does not depend on $ \brho $, but using the solution for $\bar{A}^\mu$ and $ \bar{B}^\mu$, equations \eqref{fluideq_rhoA} and \eqref{fluideq_rhoB} determine the full $ \brho $ field given the specification of an initial $ \brho $ at one point on each submanifold.

This property of forming two-dimensional submanifolds may also hold for a more general string fluid. If the tensor $ \langle F \rangle^{\mu\nu} $ annihilates exactly two linearly independent directions, it can be shown that it is a \emph{simple} bivector ---that is, there exists two vector fields $ \xi^\mu $ and $ \zeta^\mu$ such that,
 \begin{align}
 \langle F \rangle^{\mu\nu} = \xi^\mu \zeta^\nu - \zeta^\mu \xi^\nu.\label{simple}
 \end{align}
 On the other hand, the dual tensor $ \star \langle F\rangle^{\mu\nu} $ annihilates vectors in the space spanned by $ \xi^\mu $ and $ \zeta^\mu $ and, thus, the Frobenius condition for $ \xi^\mu$ and $ \zeta^\mu $ to form surfaces can be expressed as
 \begin{align}
\star \langle F\rangle_{\mu\nu}\, [\xi,\zeta]^\nu = 0.\label{forb_cond}
 \end{align}
Now if $  \langle F \rangle^{\mu\nu} $ is a simple bivector \eqref{simple}, then the conservation law
 \be
 \nabla_\mu \langle F \rangle^{\mu\nu} = (\nabla_\lambda \xi^\lambda) \zeta^\nu - (\nabla_\lambda \zeta^\lambda) \xi^\nu + [\xi,\zeta]^\nu = 0
 \ee
 which holds for any string fluid can be used to obtain the Forbenius condition \eqref{forb_cond}, 
  \begin{align}
\star \langle F\rangle_{\mu\nu}\, [\xi,\zeta]^\nu = \star \langle F\rangle_{\mu\nu}\, (- (\nabla_\lambda \xi^\lambda) \zeta^\nu + (\nabla_\lambda \zeta^\lambda) \xi^\nu ) =0.\label{frob_cond}
 \end{align}
 Once again we have used the fact that $ \star \langle F\rangle^{\mu\nu} $ annihilates vectors $ \xi^\mu $ and $ \zeta^\mu $. So under the condition of local equilibrium the fluid is foliated by a collection of submanifolds, each of which independently acts like the worldsheet of a string.

\subsection{Nambu-Goto String Dust}\label{NGDust}

 A similar ``string dust'' model was introduced by Stachel \cite{StachelDust}\cite{LetelierDust} in which each submanifold respects the Nambu-Goto action. In fact, the local equilibrium model is exactly the Stachel model when both $ \bar{A}^\mu $ and $ \bar{B}^\mu $ are restricted to be linearly independent null vectors. In that case the equations \eqref{fluideq_ADB} and \eqref{fluideq_BDA} are just the equations for a Nambu-Goto string expressed in terms of the vectors $ A^\mu$ and $ B^\mu $ defined in \eqref{AB_def} (see e.g. \cite{Transport}). Of course if there are no statistical variances, the mean $ \bar{A}^\mu $ and $ \bar{B}^\mu $ are just equal to the $ A^\mu $ and $ B^\mu $ for each individual Nambu-Goto string in the coarse-grained network, so this result would be expected.

The connection to the string dust model is more easily seen in a normalized notation. We can always choose the vectors $ \xi $ and $ \zeta $ forming $ \langle F \rangle $ in \eqref{simple} to be orthogonal, and we can also factor out any overall magnitude into a scalar $ \trho $ so that we are left with a pair of orthonormal vectors ---one timelike, $ v^\mu $, and one spacelike, $ u^\mu$, i.e.\footnote{Here the convention is that the letter $ v^\mu $ is taken to be the timelike vector, and $ u^\mu $ the spacelike vector. This notation is the opposite of the convention in certain papers, but is consistent with the notation in \cite{Fluid, Transport}.} 
\bea
v_\mu v^\mu = -u_\mu u^\mu =1\\
u_\mu v^\mu =0
\eea
 \be
 \langle F \rangle^{\mu\nu} = \trho (u^\mu v^\nu - v^\mu u^\nu).\label{simple2}
 \ee
 The unit bivector in parenthesis is denoted as
 \be
 \f^{\mu\nu} \equiv u^\mu v^\nu - v^\mu u^\nu,
 \ee
 and the quantity $ \trho $ can be found from the contraction of $ \langle F \rangle $
 \begin{align}
 \trho &\equiv \sqrt{-\frac{1}{2}\langle F\rangle^{\mu\nu} \langle F\rangle_{\mu\nu}}\label{trho}
 \end{align}
 The projector onto the submanifold $ h^{\mu\nu} $ can be also defined in terms of the unit simple bivector,
 \be
h^{\mu\nu} = \f^{\mu\rho}\f_{\rho}^{\,\,\,\nu} = v^\mu v^\nu - u^\mu u^\nu. \label{hProj}
\ee
Note that this is also the pushforward of the inverse metric $ h^{ab} $ on the worldsheet, hence the same choice of notation.

For the equilibrium string fluid the bivector magnitude is given by
\begin{align}
\trho &= \brho\sqrt{-\frac{1}{2}\bar{A}^{[\mu}\bar{B}^{\nu]} \bar{A}_{[\mu}\bar{B}_{\nu]}}\nonumber\\
	&= \frac{\brho}{2}\sqrt{(\bar{A}^{\lambda}\bar{B}_\lambda)^2 - |\bar{A}|^2|\bar{B}|^2}.
\end{align}
and if either of the variances $ \bar{A}^\mu $ or $ \bar{B}^\mu $ vanish equations \eqref{VarA} and \eqref{VarB} imply that the magnitude is proportional to the trace of the energy-momentum tensor:
\begin{align}
\trho 
 &=\frac{\brho}{2} \bar{A}^\lambda \bar{B}_\lambda\label{area_2}\\
 &=\frac{1}{2}\langle T\rangle^{\lambda}_{\,\,\lambda}.
\end{align}
Then by \eqref{area} the magnitude $ \trho $ can also be interpreted as the coarse-grained worldsheet area in the underlying string network (this will not be true when both $ \bar{A}^\mu $ and $ \bar{B}^\mu $ have statistical variance). Moreover, when both variances vanish, the simple bivector $ \langle F\rangle^{\mu\nu} $ itself can be related to $ \langle T\rangle^{\mu\nu} $,
\begin{equation}
\frac{1}{\trho}\langle F\rangle^{\mu\lambda}\langle F\rangle_{\lambda\nu} = \frac{\brho^2}{4\trho} (\bar{A}^\mu \bar{B}^\lambda \bar{A}_\lambda\bar{B}_\nu + \bar{B}^\mu \bar{B}^\lambda \bar{A}_\lambda\bar{A}_\nu)\nonumber \\
=\frac{\brho}{2} (\bar{A}^\mu \bar{B}_\nu + \bar{B}^\mu \bar{A}_\nu)\nonumber = \langle T\rangle^\mu_{\,\,\,\nu}.\label{stachel0}
\end{equation}
and using \eqref{hProj} the energy-momentum tensor may be written in terms of the bivector magnitude, $\varphi$, and unit bivector, $\f^{\mu\nu}$,
\be
\langle T\rangle^{\mu\nu} = \trho \,\f^{\mu}_{\,\,\,\lambda}\f^{\lambda\nu} = \trho h^{\mu\nu} = \trho (v^\mu v^\nu - u^\mu u^\nu).\label{stachel}
\ee

This choice of energy-momentum tensor was the starting point for the analysis in Stachel's paper \cite{StachelDust}. In our model it is seen as a special case of a coarse-grained network of strings in local equilibrium and under the condition that the statistical variations in both vectors $ {A}^\mu $ or ${B}^\mu $ are negligible.

\section{Equilibrium Fluids}\label{Sec:Solutions}

The full local equillibrium model in which there may be non-zero variances is more general than the Stachel model  \cite{StachelDust}. First consider the degenerate case in which $ \bar{A}=\bar{B} $. Then $ \langle F\rangle^{\mu\nu}$ vanishes and the energy momentum tensor becomes,
\begin{align}
\langle T\rangle^{\mu\nu} = \brho \bar{A}^\mu\bar{A}^\nu
\end{align}
which is formally equivalent to a dust of particles with four-velocity in the direction of $ \bar{A}^\mu $. In terms of the underlying string network, this represents a dust of loops which are smaller than the coarse-graining scale.

To clarify the general case when $ \bar{A}^\mu $ and $ \bar{B}^\mu $ are linearly independent, choose $ \bar{A}^\mu = (1,0,0,0) $ and $ \bar{B}^\mu = (0,1,0,0)$ to be basis vectors in the tangent space, with the other two directions orthogonal. In these coordinates, the nontrivial components of $ \langle T\rangle^{\mu}_{\,\,\,\nu}$ in equation \eqref{stachel0} can be written as the two-dimensional matrix $ \mathbb{T} $,
\begin{align}
\mathbb{T} = \frac{\brho}{2}\left(
\begin{array}{cc}
\bar{A}^\nu \bar{B}_\nu & |\bar{B}|^2 \\
|\bar{A}|^2 & \bar{A}^\nu \bar{B}_\nu
\end{array}\right).\label{matrix}
\end{align}
whose eigenvalues $\lambda$ are solutions of the characteristic equation,
 \begin{align}
\left(\frac{\brho}{2}\bar{A}^\nu \bar{B}_\nu- \lambda\right )^2 - \left(\frac{\brho}{2}\right)^2 |\bar{A}|^2 |\bar{B}|^2 = 0. \label{eigen}
 \end{align}
In a degenerate case when either $ |\bar{A}|^2 $ or $ |\bar{B}|^2 $ vanishes the only solution of \eqref{eigen} is 
 \be
 \lambda = \frac{\brho}{2} \bar{A}^\nu \bar{B}_\nu = \trho.
 \ee
 If both variances vanish the eigenspace is indeed degenerate since $ \mathbb{T} $ is just $ \trho $ multiplied by the projector on the space spanned by $ \bar{A}^\mu $ and $ \bar{B}^\mu $ ---this is just what \eqref{stachel} indicates. But if for instance $ |\bar{A}|^2 = 0 $ but $ |\bar{B}|^2 \neq 0 $, then the null vector $ \bar{A} $ is the only independent eigenvector. We will return to this case in Sec. \ref{ChiralDust}, where it will be seen that the submanifolds obey the equations of a chiral string with a null-current in the direction of $ \bar{A}^\mu $.

For now consider the case in which both $ \bar{A}^\mu $ and $ \bar{B}^\mu $ are timelike vectors. Then it is easy to verify from \eqref{matrix} that $ (\pm|\bar{A}|^{-1}, |\bar{B}|^{-1}) $ are two eigenvectors with eigenvalues $ \brho/2 (\bar{A}^\nu \bar{B}_\nu \pm |\bar{A}| |\bar{B}|) $, respectively. This suggests to renormalize $ \bar{A} $ and $ \bar{B} $ to have unit magnitude,
\begin{align}
\alpha^\mu \equiv \frac{\bar{A}^\mu}{|\bar{A}|}\nonumber\\
\beta^\mu \equiv \frac{\bar{B}^\mu}{|\bar{B}|},\label{alphabeta}
\end{align}
so that the eigenvectors are a linear combination of $ \alpha^\mu $ and $ \beta^\mu $,
\begin{align}
\V^\mu \equiv \frac{1}{2}({\beta^\mu+\alpha^\mu}) \label{V}\\
\U^\mu \equiv \frac{1}{2}({\beta^\mu-\alpha^\mu})\label{U}.
\end{align}
By the reverse Cauchy-Schwartz inequality that holds for timelike vectors, $ |\alpha^\nu \beta_\nu| \geq 1 $. This implies that $ \V^\mu $ is timelike and $ \U^\mu $ is spacelike. It is also straightforward to show that $ \V^\mu $ and $ \U^\mu $ are orthogonal 
\be
\V^\nu \U_\nu = 0
\ee
and that their magnitudes satisfy a hyperbolic relationship,
\be
|\V|^2 - |\U|^2 = 1.
\ee

\subsection{Wiggly String Dust}\label{WigglyDust}

One of the advantages to considering the normalized fields $ \alpha^\mu $ and $ \beta^\mu $ is that they have simple transformation properties. Earlier we were faced with a non-covariant rule of how to transform $ \bar{A}^\mu $ and $ \bar{B}^\mu $ under coordinate transformations. If these quantities are always defined as average propagation directions in whichever coordinates we are using then they do not transform as four-vectors. This issue can be clarified by rewriting \eqref{equilibrium} in terms of  $ \alpha^\mu $ and $ \beta^\mu $ defined in \eqref{alphabeta},
\be
\langle A\otimes B \rangle^{\mu\nu} = \brho \bar{A}^\mu \bar{B}^\nu = \prho \alpha^\mu \beta^\nu.
\ee
where 
\begin{align}
\prho \equiv \brho\,|\bar{A}|\,|\bar{B}| = \sqrt{\langle A\otimes B \rangle^{\mu\nu}\langle A\otimes B \rangle_{\mu\nu}}
\end{align}
is a scalar quantity and  $ \alpha^\mu $ and $ \beta^\mu $ are the unit four-vectors and thus transform covariantly under coordinate transformations. 

In terms of the newly defined quantities the fluid equations \eqref{AB_mu} and \eqref{AB_nu} can be rewritten in manifestly covariant form,
\begin{align}
\nabla_\mu ({\prho} \alpha^\mu\beta^\nu) = 0\label{AB_mu_2}\\
\nabla_\nu ({\prho} \alpha^\mu\beta^\nu) = 0\label{AB_nu_2}.
\end{align}
As before, we can decouple the equations by contracting \eqref{AB_mu_2} and \eqref{AB_nu_2}  with $ \beta_\nu $ and $ \alpha_\nu$ respectively
\begin{align}
\beta_\nu \nabla_\mu ({\prho} \alpha^\mu\beta^\nu) = \beta_\nu (\beta^\nu\nabla_\mu (\prho \alpha^\mu) + \prho \alpha^\mu\nabla_\mu \beta^\nu) = 0\\
\alpha_\nu \nabla_\nu ({\prho} \alpha^\mu\beta^\nu) = \alpha_\nu (\alpha^\nu\nabla_\mu (\prho \beta^\mu) + \prho \beta^\mu\nabla_\mu \alpha^\nu) = 0.
\end{align}
Then using the normalization conditions
\be
\alpha_\mu \alpha^\mu = \beta_\mu \beta^\mu =1
\ee
and
\be
\beta_\nu \nabla_\mu \beta^\nu=\alpha_\nu \nabla_\mu \alpha^\nu = 0
\ee
we recover two equations,
\begin{align}
\nabla_\mu (\prho \alpha^\mu)  = 0 \label{fluideq_rhoAlpha}\\
\nabla_\mu (\prho \beta^\mu) = 0\label{fluideq_rhoBeta}
\end{align}
which can be substituted back into \eqref{AB_mu_2} and \eqref{AB_nu_2} to obtain two more equations
\begin{align}
\alpha^\mu\nabla_\mu \beta^\nu = 0,\label{fluideq_ADBeta} \\
\beta^\mu\nabla_\mu \alpha^\nu = 0.\label{fluideq_BDAlpha}
\end{align}
Note that equations \eqref{fluideq_ADBeta} and \eqref{fluideq_BDAlpha} imply that $\alpha^\mu$ and $\beta^\mu$ are the basis vectors for some coordinates on the submanifolds since
\be
[\alpha, \beta]^\nu =  \alpha^\mu\nabla_\mu \beta^\nu - \beta^\mu\nabla_\mu \alpha^\nu =0.
\ee 
Moreover the scalar $ \prho $ had completely decoupled from these equations and is determined by equations \eqref{fluideq_rhoAlpha} and \eqref{fluideq_rhoBeta}. 

The equations \eqref{fluideq_ADBeta} and \eqref{fluideq_BDAlpha}  may also be rewritten in terms of the eigenvectors $ U^\mu $ and $ V^\mu $ related to $\alpha^\mu$ and $\beta^\mu$ through equations \eqref{V} and \eqref{U}, 
\begin{align}
V^\mu \nabla_\mu U^\nu - U^\mu \nabla_\mu V^\nu = 0\label{fluideq_uvFrob}\\
V^\mu \nabla_\mu V^\nu - U^\mu \nabla_\mu U^\nu = 0\label{fluideq_uvWave}.
\end{align}
The vanishing of the commutator of $ U $ and $ V $ in \eqref{fluideq_uvFrob} indicates that $U^\mu$ and $V^\mu$ are also coordinate basis vectors for some coordinates $ \sigma $ and $ \tau $ on a submanifold, i.e.
\begin{align}
V^\mu = \frac{\partial X^\mu}{\partial \tau}\nonumber\\
U^\mu =  \frac{\partial X^\mu}{\partial \sigma}.
\end{align}
Then equation \eqref{fluideq_uvWave} can be view as a wave equation for the embedding of the submanifold coordinates in the target space. For example, in flat spacetime equation \eqref{fluideq_uvWave} reduces to
\begin{align}
\frac{\partial^2 X^\mu}{\partial\tau^2}-\frac{\partial^2 X^\mu}{\partial\sigma^2}=0.\label{fluideq_x}
\end{align}
where in contrast to the Nambu-Goto case the coordinates $ \sigma $ and $ \tau $ are not necessarily conformal. Instead these equations for the submanifold are equivalent to those of a wiggly string. The wave equation \eqref{fluideq_x} appears in terms of timelike $ \bar{A}^\mu $ and $ \bar{B}^\mu $ in a paper by Vilenkin \cite{VilenkinWiggly}, and the equations \eqref{fluideq_ADBeta} and \eqref{fluideq_BDAlpha} for $ \alpha^\mu $ and $ \beta^\mu $  appear in a paper by Carter \cite{CarterWiggly}.

To further see that the submanifold obeys the wiggly string equation of state, notice that \eqref{fluideq_uvWave} can be interpreted as the conservation of a tensor current on the submanifold, much like the Nambu-Goto equation \eqref{T_current} was related to the conservation of the energy-momentum tensor \eqref{T}. Similarly to equation \eqref{T} we can define a conserved but singular energy-momentum tensor 
\be
T^{\mu\nu}(x^\lambda) \sqrt{-g(x^\lambda)}= \int d \eta_0 \wedge d\eta_1\,  \tilde{T}^{\mu\nu}(\eta)\,  \delta(x^\lambda-X^\lambda)
\ee
with support on the submanifold, which involves the pushforward of a worldsheet current to the target space,
\begin{align}
\tilde{T}^{\mu\nu} &= \V^\mu\V^\nu - \U^\mu\U^\nu.\label{T_tilde}
\end{align}
The main difference now is that $ \tilde{T}^{\mu\nu} $ can be defined for quite general models of strings in terms of the surface energy density $ M $ and the surface tension $ T $ \cite{CarterBrane},
\begin{align}
\tilde{T}^{\mu\nu} &= \sqrt{-h}(M v^\mu v^\nu - T u^\mu u^\nu),\label{T_MT}
\end{align}
where as before $ v^\mu$ and $ u^\mu $ are the unit eigenvectors of the energy-momentum tensor. But in the $ \sigma, \tau $ coordinate system the induced metric \eqref{hmetric} is 
\be
h_{ab} = \left(
\begin{array}{cc}
 V^\mu V_\mu   &  V^\mu U_\mu \\
 U^\mu V_\mu   &  U^\mu U_\mu 
\end{array}\right)
\ee
and thus
\be
\sqrt{-h} = |\V|\,|\U|.
\ee
Then equations \eqref{T_tilde} and \eqref{T_tilde} imply,
\begin{align}
M &= \frac{|V|}{|U|}\\
T &= \frac{|U|}{|V|}\label{MTUV}
\end{align}
and the submanifold indeed obey the wiggly string equation of state \cite{CarterWiggly}:
\begin{align}
M\,T = 1.
\end{align}

\subsection {Chiral String Dust}\label{ChiralDust}

Now we come back to the remaining case when the statistical variance in only one of the propagating directions vanishes. Without loss of generality we can assume that the corse-grained tensors
\be
\langle A \otimes B \rangle^{\mu\nu} = \brho \bar{A}^\mu \bar{B}^\nu \label{AB_chiral}
\ee
where $A^\mu$ is a null vector (or $ |\bar{A}| = 0$) and $B^\mu$ is a time-like vector (or $ |\bar{B}|>0$). In flat spacetime the equations of motion \eqref{fluideq_ADB} and \eqref{fluideq_BDA} reduce to the wave equation \eqref{fluideq_x} with the difference that the spatial part of $\bar{B} $ lies inside of the Kibble-Turok sphere. This is just the equation of motion for a chiral string \cite{CarterChiral, KibbleChiral, VilenkinChiral}. We will further show that the submanifolds obey the equations of a chiral string in arbitrary background metric.

We can renormalize $ \brho $ to the scalar $\trho$ defined by \eqref{trho} and given by \eqref{area_2}, $ \bar{B}^\mu $ to a unit vector $ \beta^\mu $, and then absorb all of the normalization factors into a new vector $ n^\mu $ in the direction of $ \bar{A}^\mu $,
\begin{align}
\trho &= \frac{\brho}{2} (\bar{A}^\lambda\bar{B}_\lambda)\\
\beta^\mu &= \frac{B^\mu}{ |\bar{B}|}\\
n^\mu &\equiv \frac {2 |\bar{B}| \bar{A}^\mu}{\bar{A}^\lambda\bar{B}_\lambda}
\end{align}
so that \eqref{AB_chiral} can be written as
\be
\langle A \otimes B \rangle^{\mu\nu} = \trho \beta^\mu n^\nu.
\ee
Following the Carter and Peter's paper on the chiral string model \cite{CarterChiral} we can define the other linearly independent null vector,
\begin{align}
m^\mu \equiv  \beta^\mu - \frac{1}{2}  n^\mu,
\end{align}
then 
\begin{align}
m^\mu m_\mu=\beta^\mu \beta_\mu - \frac{1}{2} \beta^\mu n_\mu + \frac{1}{4} n^\mu n_\mu  = 0 \\
m^\mu n_\mu= \beta^\mu n_\mu - \frac{1}{2} \beta^\mu n_\mu  = 1.
\end{align}

By considering the conservation equations for $\langle A \otimes B \rangle^{\mu\nu}$ in the same manner as before we find:
\begin{align}
\nabla_\lambda(\trho n^\lambda) = 0 \label{chiral1} \\
n^\lambda \nabla_\lambda \beta^\mu = 0. \label{chiral2}
\end{align}
and by contracting \eqref{chiral2} with $2 m_\mu $ we see that $ n^\mu $ is indeed a conserved null current.
\be
2 m_{\mu}n^\lambda \nabla_\lambda \beta^\mu = m_{\mu}n^\lambda \nabla_\lambda (2 m^\mu + n^\mu)=  (m_{\mu}n^\lambda + n_{\mu}m^\lambda  )\nabla_\lambda n^\mu =0
\ee
or
\be
h^\lambda_{\,\,\,\mu}\nabla_\lambda n^\mu = 0.
\ee
This can be also written as
\be
h^{\lambda\mu} = n^{(\lambda} m^{\mu)} 
\ee is a projector on the worldsheet as in equation \eqref{hProj}.  Taking the surface energy-momentum tensor $ T^{\mu\nu}$ as usual to be $ \langle T \rangle^{\mu\nu} $ with $ \trho $ factored out,
\begin{align}
T^{\lambda\mu} &= n^{(\lambda}\beta^{\mu)}\\
				&= n^{\lambda}n^\mu + n^{(\lambda}m^{\mu)} = n^{\lambda}n^\mu + h^{\lambda\mu},
\end{align}
which again agrees with the chiral string model in \cite{CarterChiral,KibbleChiral,VilenkinChiral}.

\section{Discussion}\label{Sec:Discussion}

In this paper we studied the solutions of the string fluids equation under assumption of local equilibrium. Although the true equilibrium is never established the local equilibrium assumption is often a starting point for analyzing the behavior of the fluid equations. A distinguishing feature of the equilibrium fluids is that the space-time can be foliated into non-interacting two-dimentional submanifolds. Then the equations of motion describing the submanifolds can be used to identify and classify the three different classes of string fluids: Nambu-Goto string dust, chiral string dust and wiggly string dust. These fluids are described respectively by $4$, $5$ and $6$ degrees of freedom, where $6$ is the largest number of dynamical equations which can be obtained from the conservation equations.\footnote{In addition there is the freedom to specify the energy density at one point on each submanifold.}

The Nambu-Goto string dust corresponds to the least generic case in which the statistical variance of both the right and left-moving null directions vanishes. This solution describes a dust of Nambu-Goto strings first studied by Statchel \cite{StachelDust} in a different context. In Sec.  \ref{NGDust} we described an explicit connection to the Stachel model using slightly different notations. 

The second class corresponds to an equilibrium fluid of strings with submanifolds described by the action of chiral strings first proposed by Witten \cite{WittenChiral}. In terms of string fluids it describes the submanifolds of a chiral string dust discussed in Sec. \ref{ChiralDust} where the known results about chiral strings were derived from the fluid equations. In the model of chiral string dust only one of the statistical variances either in the left- or right-moving null directions is negligible.  

The most general class of solutions of the equilibrium fluids corresponds to the wiggly string dust described in Sec. \ref{WigglyDust}. The submanifolds of these fluids are given by the equations of motion of the wiggly strings considered previously by Vilenkin \cite{VilenkinWiggly} and Carter \cite{CarterWiggly}. In the wiggly string dust model the variances of both the left- and right-moving direction do not vanish as one would generically expect. 

The equilibrium string fluid is quite general in a sense that it can simultaneously describe the different types of strings such as Nambu-Goto strings, chiral strings and wiggly strings, but may not be general enough to describe the networks of strings phenomenologically. For example, it is known that the intersections of the submanifolds of the equilibrium fluids would generically lead to shock waves that can only be resolved by higher order terms \cite{Fluid}. The inclusion of such terms would be essential in order to describe the networks of, for example, cosmic strings using string fluids. 

Consider the following phenomenological expansion of the spatial components of the $\langle A \otimes B \rangle^{\mu\nu}$ tensor,
\be
\langle A\otimes B\rangle^{i j} = \rho \bar{A}^i \bar{B}^j + \frac{w}{2} \rho g^{i j} - \frac{\alpha}{4} \left (\partial^{(i}  \rho \bar{A}^{j)} +  \partial^{(i} \rho \bar{B}^{j)} \right ) + \frac{\beta}{4} \left ( \partial^{[i} \rho \bar{A}^{j]} + \partial^{[j} \rho \bar{B}^{i]} \right ) + ...,
\ee
where $w$ and $\alpha$ are the equation of state parameter and viscous coefficients of a Newtonian fluid, but $\beta$ describes non-Newtonian viscous effects in the string fluid. Then the transport coefficients $w$, $\alpha$ and $\beta$ can be extracted directly from numerical simulations of the Nambu-Goto strings \cite{NumericalFlat, NumericalLoops, NumericalFriedmann, NumericalParallel} or obtained analytically from the kinetic theory of strings \cite{Kinetic, Transport}. The analytical and numerical analysis of the non-equilibrium transport phenomena will be the subject of the upcoming publication \cite{ViscousFluid}.

\end{document}